\documentclass[5p]{elsarticle}
\usepackage{graphicx}
\usepackage[T1]{fontenc}
\usepackage{amsmath}
\usepackage{hyperref}
\usepackage{float}

\begin{document}

\title{Energy decomposition analysis of neutral and negatively charged borophenes}

\author{T. Tarkowski}
%\ead{tomasz.tarkowski@fuw.edu.pl}

\author{J. A. Majewski}
%\ead{jacek.majewski@fuw.edu.pl}

\author{N. Gonzalez Szwacki\corref{cor}}
\ead{gonz@fuw.edu.pl}
\cortext[cor]{Corresponding author.}

\address{Institute of Theoretical Physics, Faculty of Physics, University of Warsaw, Pasteura~5, PL-02093 Warszawa, Poland}

\begin{abstract}
The effect of external static charging on borophenes -- 2D boron crystals -- is investigated by using first principles calculations. The influence of the excess negative charge on the stability of the 2D structures is examined using a very simple analysis of decomposition of the binding energy of a given boron layer into contributions coming from boron atoms that have different coordination numbers. This analysis is important to understand how the local neighbourhood of an atom influences the overall stability of the monolayer structure. The decomposition is done for the $\alpha$-sheet and its related family of structures. From this analysis, we have found a preference for 2D boron crystals with very small or very high charges per atom. The structures with intermediate charges are energetically not favourable. We have also found a clear preference in terms of binding energy for the experimentally seen $\gamma$-sheet and $\delta$-sheet structures that is almost independent on the considered excess of negative charge of the structures. On the other hand, we have shown that a model based solely on nearest-neighbour interactions, although instructive, is too simple to predict binding energies accurately.
\end{abstract}

\begin{keyword}
\textit{ab initio} calculations \sep borophene \sep energy decomposition

\PACS 31.15.A- \sep 31.15.es \sep 61.46.-w \sep 62.23.Kn \sep 62.25.-g \sep 68.65.-k
\end{keyword}

% 31.15.A- 	Ab initio calculations
% 31.15.es 	Applications of density-functional theory [...]
% 61.46.-w 	Structure of nanoscale materials [...]
% 62.23.Kn 	Nanosheets
% 62.25.-g 	Mechanical properties of nanoscale systems [...]
% 68.65.-k 	Low-dimensional, mesoscopic, nanoscale and other related systems: structure and nonelectronic properties [...]

\maketitle 

\section{Introduction}

Like carbon, boron can adopt bonding configurations that favor the formation of low-dimensional structures such as nanotubes, fullerenes, and sheets. With these different forms (or allotropes) could come interesting and novel properties distinct from those of the bulk structures. Two-dimensional (2D) allotropes of boron have been studied theoretically more extensively during the last ten years \cite{li2015}. There are different proposals based on first principles calculations for the atomic structure of 2D boron crystals: buckled triangular (bt) sheet \cite{evans2005,kunstmann2006,cabria2006}, $\alpha$-sheet and related planar and quasiplanar layers \cite{tang2007,penev2012,yu2012,wu2012}, non-zero thick layers and bilayers \cite{lau2007,zhou2014,ma2016}, B$_{12}$-based layers \cite{lau2007}, and other structures \cite{ngsz2007,yu2012}. However, the experimental realization came just recently and confirmed only some of those structures.

\begin{figure}[H]
\centering
  \begin{tabular}{ll}
    (a) $s_3$ (hc) & $s_4$ \\
    \includegraphics[width=.45\linewidth]{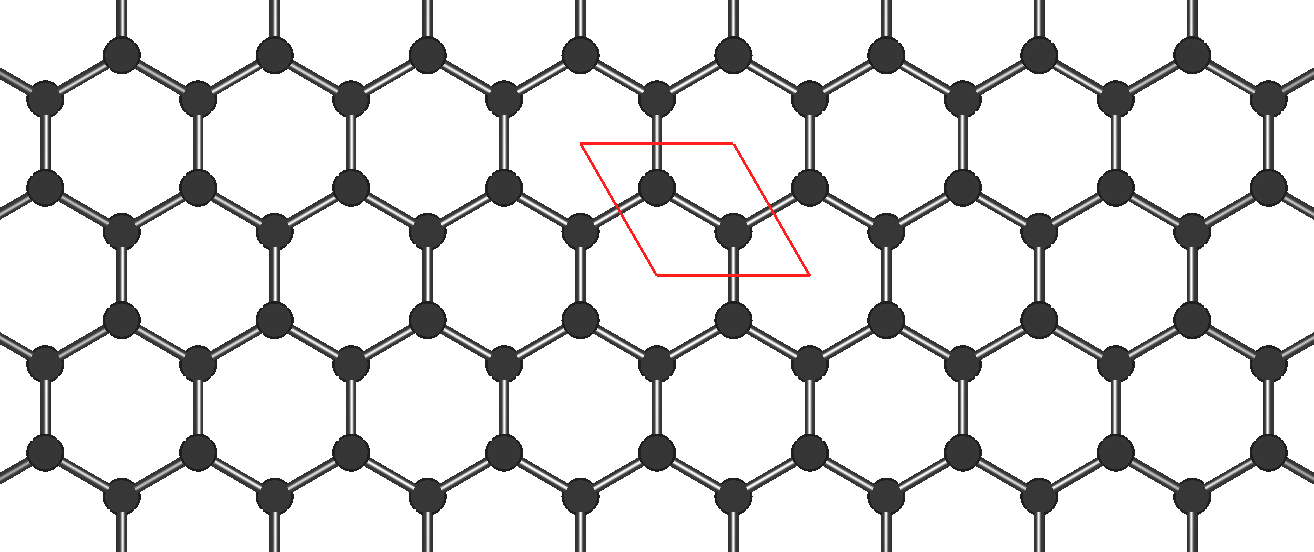} & \includegraphics[width=.45\linewidth]{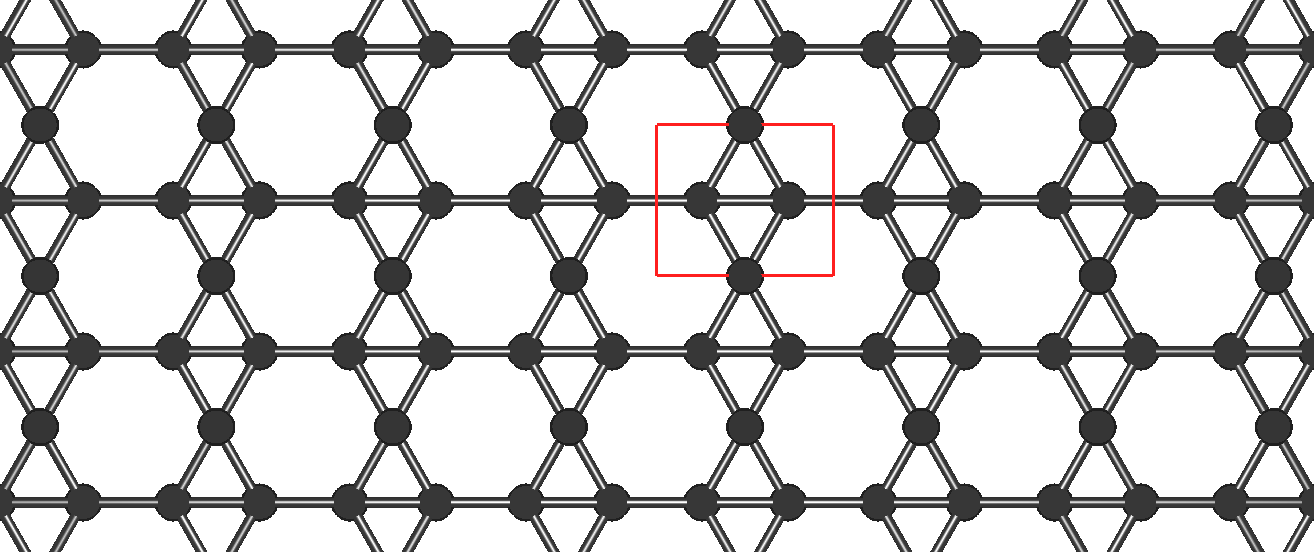} \\
    $s_5$ & $s_6$ (bt) \\
    \includegraphics[width=.45\linewidth]{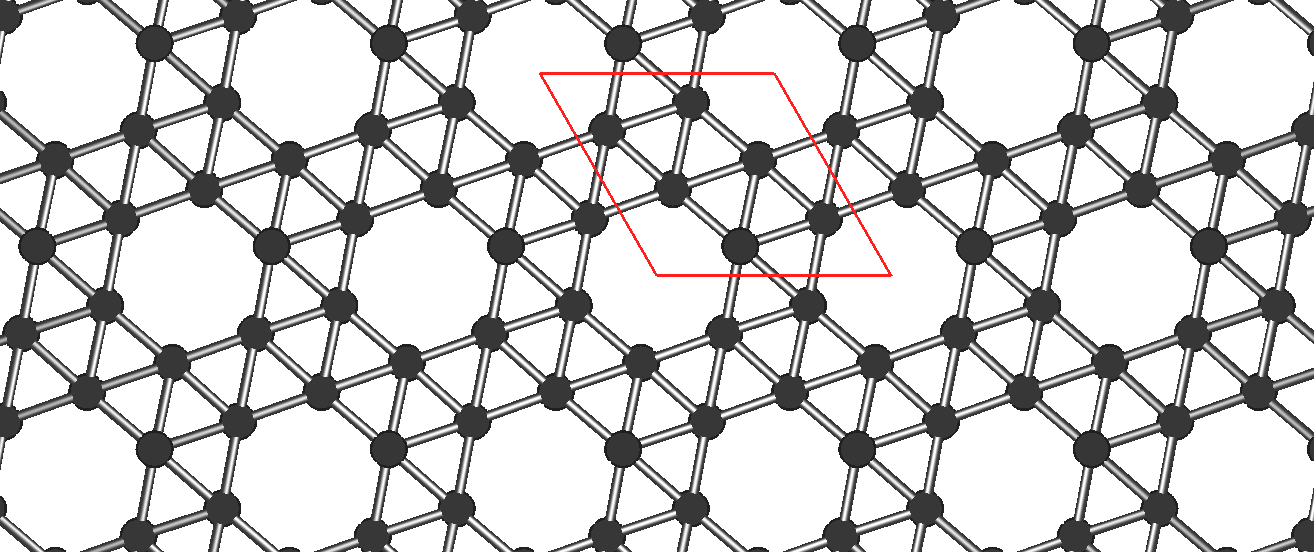} & \includegraphics[width=.45\linewidth]{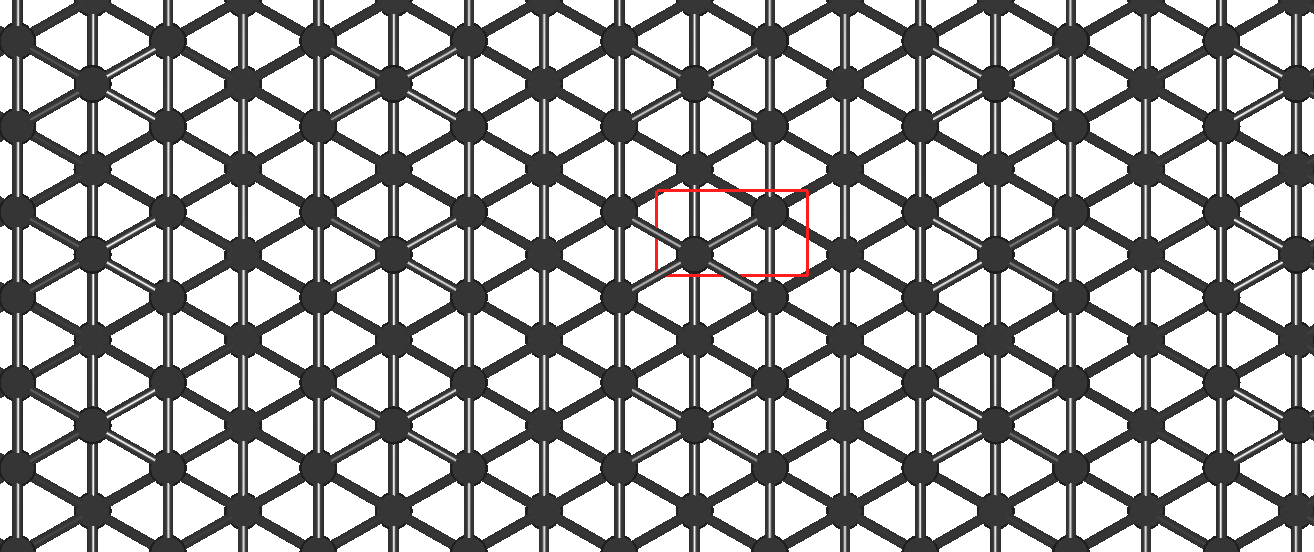} \\
    (b) $\alpha$-sheet  & $\beta$-sheet \\
    \includegraphics[width=.45\linewidth]{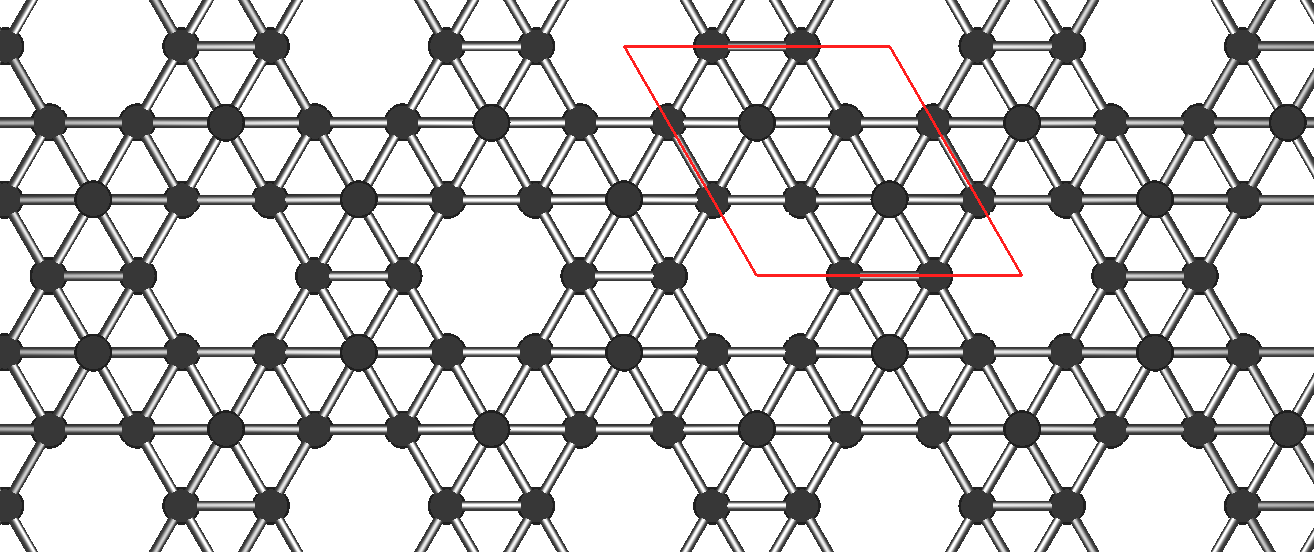} & \includegraphics[width=.45\linewidth]{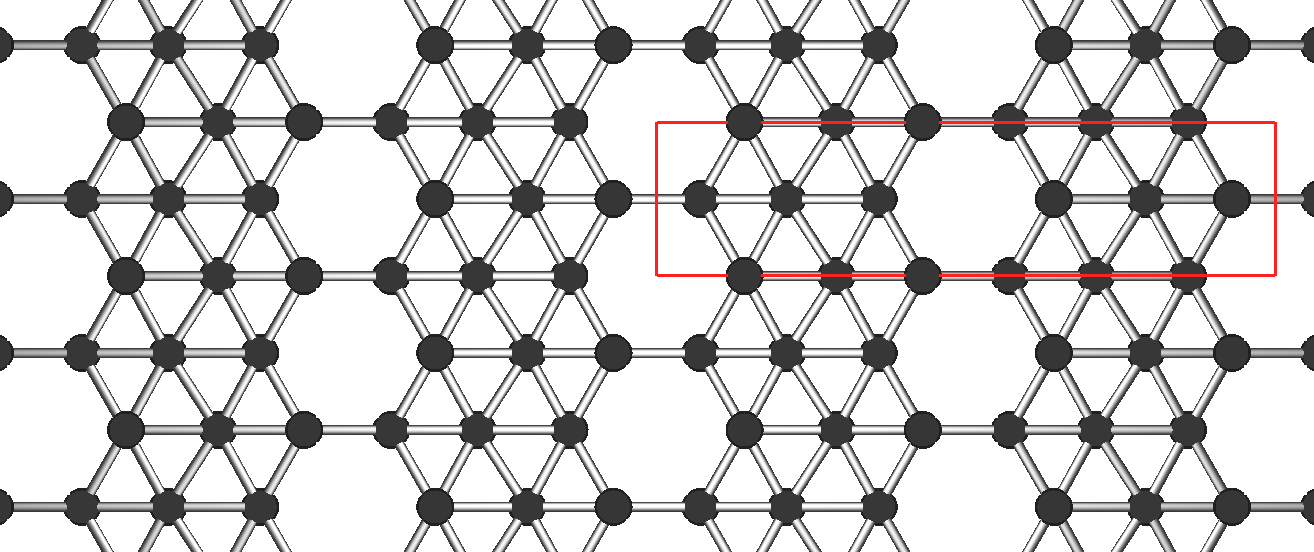} \\
    $\gamma$-sheet  & $\delta$-sheet \\
    \includegraphics[width=.45\linewidth]{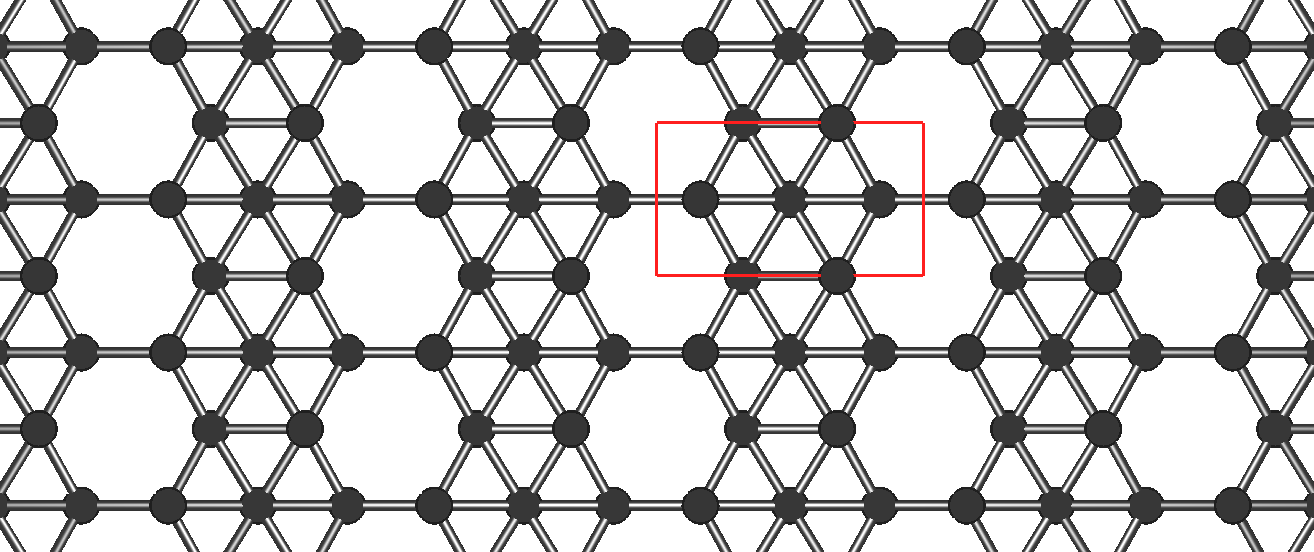} & \includegraphics[width=.45\linewidth]{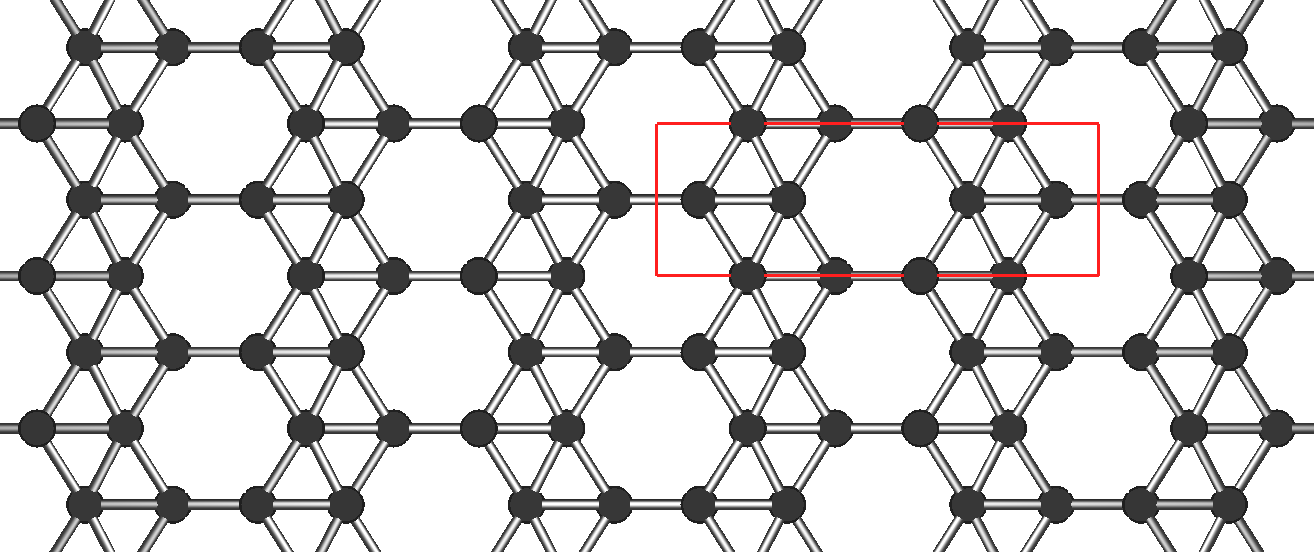} \\
    \multicolumn{2}{c}{$\varepsilon$-sheet} \\
    \multicolumn{2}{c}{\includegraphics[width=.45\linewidth]{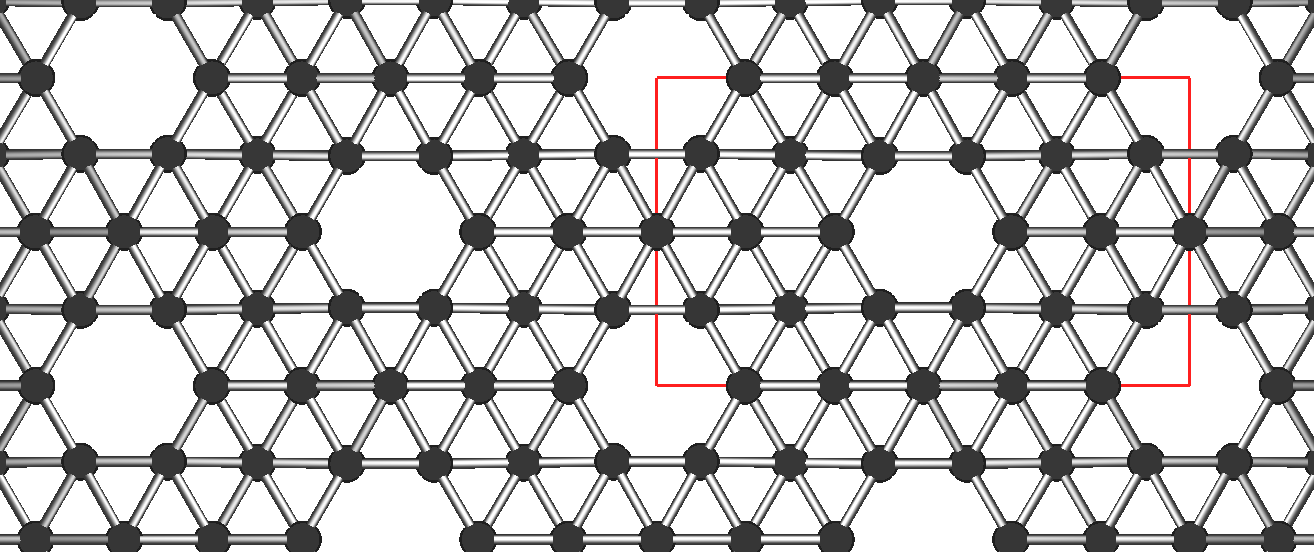}} \\
  \end{tabular}
  \caption{(Color online). 2D boron layers considered in this study. (a) Structures that are used to calculate the values of the $e_i$ energies in our model. (b) Structures that are used to test the accuracy of our model. In all cases, the conventional unit cells are shown in red.}
\label{fig1}
\end{figure}

G. Tai \textit{et al.} \cite{tai2015} reported the synthesis of 2D boron structures on copper foils by chemical vapor deposition (CVD), by combining boron and B$_{2}$O$_{3}$ to make B$_{2}$O$_{2}$ vapour and reducing it with hydrogen while passing it over copper foil. Their 2D boron structure consists of B$_{12}$ icosahedra held together by B$_{2}$ dumbbells and behaves as a direct band gap semiconductor. The same year, Mannix \textit{et al.} \cite{mannix2015} reported the synthesis of 2D boron sheets grown on a single crystal Ag(111) substrate using molecular beam epitaxy (MBE) under ultrahigh-vacuum (UHV) conditions. Two distinct forms of 2D boron structures have been observed, both consisting of triangular layers with some fraction of missing atoms (empty hexagons, hexagonal holes or vacancies) in the hexagonal lattice. In one form (labeled in this work as $\gamma$-sheet, see Fig.~\ref{fig1}b) rows of filled hexagons are separated by chains of empty hexagons; in the other (labeled in this work as $\delta$-sheet, see Fig.~\ref{fig1}b), boron atoms take up narrow zigzag stripes separated by arrays of empty hexagons. Should be noted that the first form was initially proposed to be the bt sheet (see Fig.~\ref{fig1}a) \cite{mannix2015}, and then recently confirmed to be the $\gamma$-sheet \cite{zhang2016_2}. Similarly, Feng \textit{et al.} \cite{feng2016} used MBE to grow 2D sheets of boron on a metallic Ag(111) substrate by direct evaporation of a pure boron source under UHV conditions. They also observed two different structures corresponding to the described above $\gamma$ and $\delta$ sheets. According to this study, both sheets are flat, metallic in character, and quite stable against oxidation in air. Moreover, the sheets appear to be robust and only weakly bound to their substrate, indicating that it might be possible to obtain freestanding sheets, but the question of how to detach the sheets from the substrates is still open. Finally, in the most recent experimental study \cite{zhong2017} two reproducible metallic phases of 2D boron are found on Ag(111), one of them being the $\gamma$-sheet and the other one presumably being the $\alpha$-sheet. All these experimental studies pave the way to fascinating applications in nanoelectronic and nanophotonic devices \cite{li2015,tai2015,zhang2016}.

There are still many unanswered questions about the structure and properties of 2D boron structures. In this work, we concentrate only on monolayer boron structures. For these structures it is not clear, for instance, why from the vast number of distinct 2D boron sheets reported theoretically \cite{zhang2016}, only some of them seem to be favored and realized experimentally. Also, the high-symmetry $\alpha$-sheet that has been predicted to be one of the most stable one-atom-thick forms of 2D boron is believed to be obtained just very recently \cite{zhong2017}. As of our understanding, the main factors that may influence the structure of 2D boron crystals are the strain induced by the substrate and the amount of negative charge that is transferring from the substrate \cite{zhang2016}. In this study, we propose a simple model to predict the structure of 2D boron crystals exposed to static negative charge. The model does not include the strain but may serve as a first step in the search for stable charged boron sheets.

\section{Computational approach}
Our first principles calculations are based on density functional theory (DFT) and the projector augmented wave (PAW) method as implemented in the \textsc{Quantum ES\-PRES\-SO} simulation package \cite{qe2009}. For the exchange and correlation functional, we use a revised Perdew-Burke-Ern\-zer\-hof spin-polarized generalized gradient approximation (PBEsol-GGA) functional. The plane-wave basis set is converged using a 60~Ry energy cutoff. A $8\times 8\times 1$ \textbf{\textit{k}}-point mesh and a Gaussian smearing of 0.005~Ry is used in the Brillouin Zone integration. The calculations are done using supercells ensuring a 50~{\AA} separation between adjacent layers. For the charged structures, the amount of negative charge (excess of electrons) is specified in units of the charge of an electron per boron atom. For each considered structure, we do a full atomic position and lattice parameter relaxation. The 2D bulk modulus for the hc-sheet ($s_3$) is obtained from the Murnaghan equation of state. 

Images of the crystal structures shown in Fig.~\ref{fig1} were created using the VESTA visualization program \cite{VESTA}. The space group symmetries are found by using the FINDSYM software package \cite{findsym} and later on reduced to plane group symmetries. The \textbf{\textit{k}}-point mesh used in the density of states (DOS) and charge density post-processing calculations was $10\times 10\times 1$.

\section{Results and discussion}
\subsection{Static charging of 2D boron structures}

\begin{figure}
\centering
\includegraphics[width=.99\linewidth]{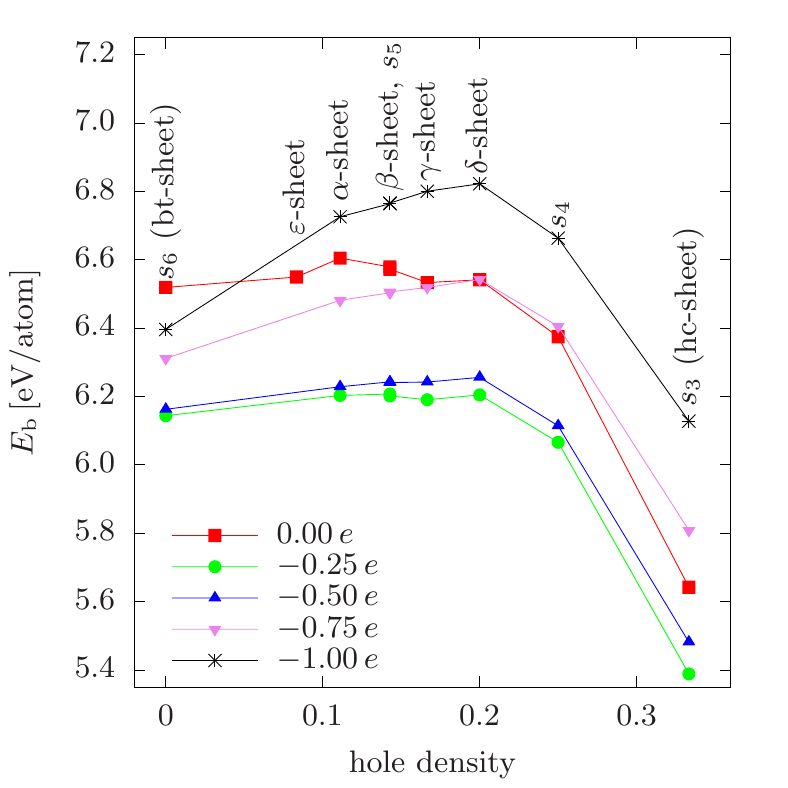}
\caption{(Color online). Binding energy versus hole density for neutral and negatively charged boron sheets. The two limiting cases of $0$ and $1/3$ hole density correspond to the buckled triangular and honeycomb sheets, respectively. Maximum $E_\textnormal{b}$ occurs for sheets $\alpha$ and $\delta$ for neutral and charged structures, respectively. Two structures, $\beta$-sheet and $s_5$, have the same hole density values and a slightly different (by few meV) binding energy per atom.}
\label{fig2}
\end{figure}

The structure of all the theoretically and experimentally reported one-atom-thick 2D boron crystals can be easily compared to that of graphene, with that difference that part of the hexagons in the boron layers are filled with additional boron atoms. Should be pointed out that the $\alpha$, $\gamma$, and $\delta$ sheets with periodic arrangements of empty hexagons have been extensively studied theoretically either as freestanding structures or on metallic and nonmetallic substrates long before the actual experimental realization of these structures \cite{ozdogan2010,amsler2013,liu2013,liu2013_2,zhang2015}. In the extreme case of all the hexagons filled with atoms, we get the bt-sheet mentioned above that was extensively studied theoretically at the earliest stage of the 2D boron investigation \cite{evans2005,kunstmann2006,cabria2006}. On the other hand, the boron sheet with honeycomb (hc) structure is unstable with respect to shearing perturbations \cite{evans2005}. However, for the purpose of our analysis, we fix the symmetry of the boron hc-sheet during relaxation in order to preserve its structure. We define the hexagon hole density of the boron layer in the same way as in Ref.~\cite{tang2007}, namely as the ratio between the number of missing B atoms in the triangular sheet and the number of atoms in the fully filled sheet. In that way, the bt and hc sheets have hexagon hole densities equal to $0$ and $1/3$, respectively. The full set of structures, with different hole densities considered in this work, is shown in Fig.~\ref{fig1}.

\begin{figure}
\centering
\includegraphics[width=.99\linewidth]{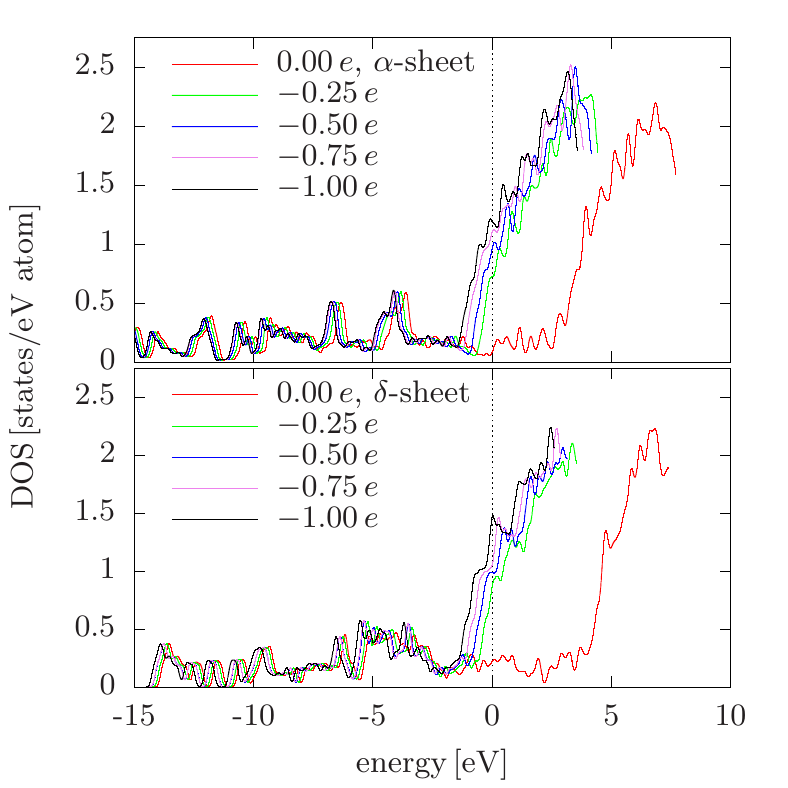}
\caption{(Color online). Density of states for neutral and negatively charged $\alpha$ (top) and $\delta$ (bottom) sheets. The zero of energy is set to the Fermi level.}
\label{fig5}
\end{figure}

In Fig.~\ref{fig2}, we have plotted the dependence of the binding energy, $E_\textnormal{b}=E_{\textnormal{tot}}(\textnormal{isolated B atom})-E_{\textnormal{tot}}(\textnormal{B layer})/N$, where $E_{\textnormal{tot}}$ is the computed total energy and $N$ is the number of atoms in the supercell, versus the hexagon hole density for structures with different static charges $q$. For neutral structures, the picture is the same as reported in Ref.~\cite{tang2007}, however, it is clear from this figure that static negative charge gives preference in terms of energy to the $\delta$-sheet and next, for $|q|>0.25\ e/\textnormal{atom}$, to the $\gamma$-sheet, that is, to structures that were recently reported experimentally \cite{feng2016}. Interestingly enough, almost all 2D layers studied here have the smallest and the largest $E_\textnormal{b}$ values for $q=-0.25\ e/\textnormal{atom}$ and $q=-1.00\ e/\textnormal{atom}$, respectively. The exception is the bt-sheet that has the largest $E_\textnormal{b}$ for the neutral structure. In Fig.~\ref{fig2}, we can also see that for $|q|\le0.5\ e/\textnormal{atom}$, there is a small dependence of $E_\textnormal{b}$ on the hexagon hole density for structures with hole densities ranging from $0$ to $0.2$.  On the other hand, for the highly charged case of $q=-1.00\ e/\textnormal{atom}$, there is a well defined maximum in $E_\textnormal{b}$ that corresponds to the $\delta$-sheet.

\begin{figure}
\centering
\includegraphics[width=.99\linewidth]{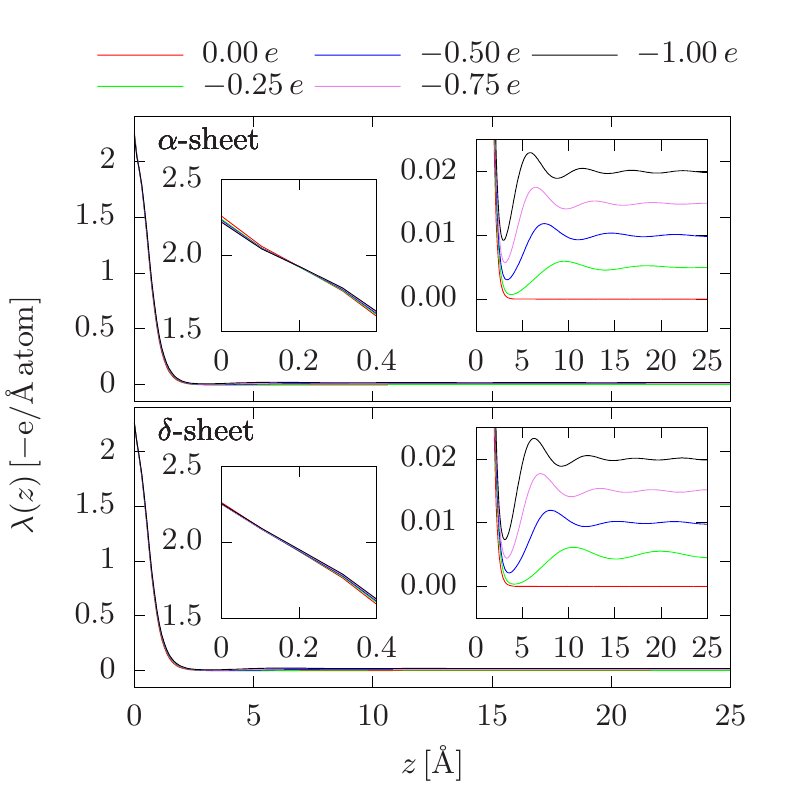}
\caption{(Color online).  Planarly averaged charge density $\lambda(z) = \frac{1 }{N} \int_{V_{2\textnormal{D}}}  \frac{\partial^3 Q }{\partial x \partial y \partial z} dxdy$, where $N$ is the number of atoms in the unit cell of volume $V_{2\textnormal{D}}$, for neutral and negatively charged $\alpha$ (top) and $\delta$ (bottom) sheets. The insets show close-ups of the relevant regions of the charge density distributions.}
\label{fig6}
\end{figure}

The DOS for the neutral and charged $\alpha$ and $\delta$ sheets is plotted in Fig.~\ref{fig5}. For the neutral structures, it is clear that the $\alpha$-sheet has a slightly smaller DOS at the Fermi level making this structure, in principle, more stable but less conductive. Our DOS calculations and also those reported in the literature \cite{ozdogan2010,penev2012,yu2012} predict a metallic behavior for the $\gamma$-sheet and all the other considered in this work 2D structures. Should be mentioned, however, that calculations using the PBE0 hybrid functional anticipate that the $\alpha$-sheet is a semiconductor \cite{wu2012}. In the case of the charged boron sheets, the excess of negative charge populates the surface states even for the smallest considered negative charge. This is shown in Fig.~\ref{fig5} top and bottom on the example of the $\alpha$ and $\delta$ sheets, respectively. A similar picture was reported for static charging of graphene \cite{topsakal2011} for which the surface states get occupied with electrons for charges as small as $q=-0.0115\ e/\textnormal{atom}$.

The real-space planarly averaged valence charge density distribution for the neutral and charged $\alpha$ and $\delta$ sheets is shown in Fig.~\ref{fig6} top and bottom, respectively. As can be seen in the figure, the charge distributions for those two structures are almost identical (see the insets of the figure). For the neutral structures, the charge density is nearly entirely localized within $\sim2.5~\textnormal{\AA}$ at both sides from the layers. The relevant portion of the excess of negative charge is distributed from $\sim2.5~\textnormal{\AA}$ to about $15~\textnormal{\AA}$ from the layers (see the right insets of Fig.~\ref{fig6}). It is interesting to note that the negative charge gets closer towards the planar structures with the increase of excess of charge. For each  considered case, there is also a small portion of negative charge distributed practically evenly across the vacuum region of the cells, what is an artifact of our periodic DFT calculations. A similar picture extends also for the rest of the considered structures.

\subsection{Energy decomposition analysis}

\begin{figure}
\centering
  \includegraphics[width=.99\linewidth]{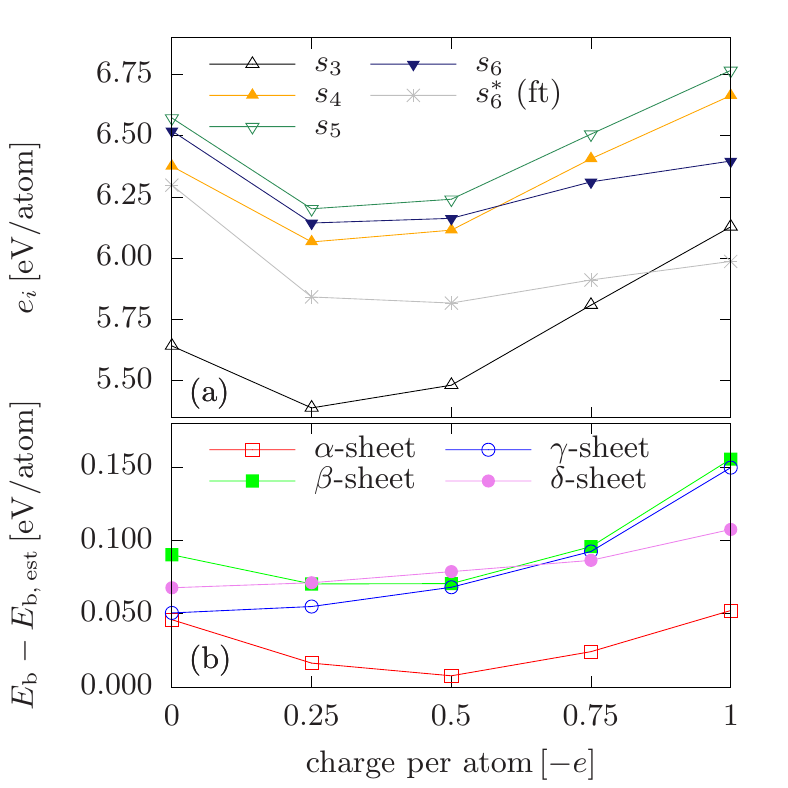}
  \caption{(Color online). (a)~Individual energy contributions, $e_i$, used in our model versus the excess of negative charge. The $e_i$ energies are binding energies per atom of structures with only one type of coordination number $i$, shown in Fig.~\ref{fig1}a. (b)~Difference between calculated, $E_{\textnormal{b}}$, and  estimated (from the model), $E_{\textnormal{b, est}}$, binding energies plotted as a function of the excess of negative charge for the structures shown in Fig.~\ref{fig1}b.}
\label{fig3}
\end{figure}

\begin{figure}
\centering
  \includegraphics[width=.99\linewidth]{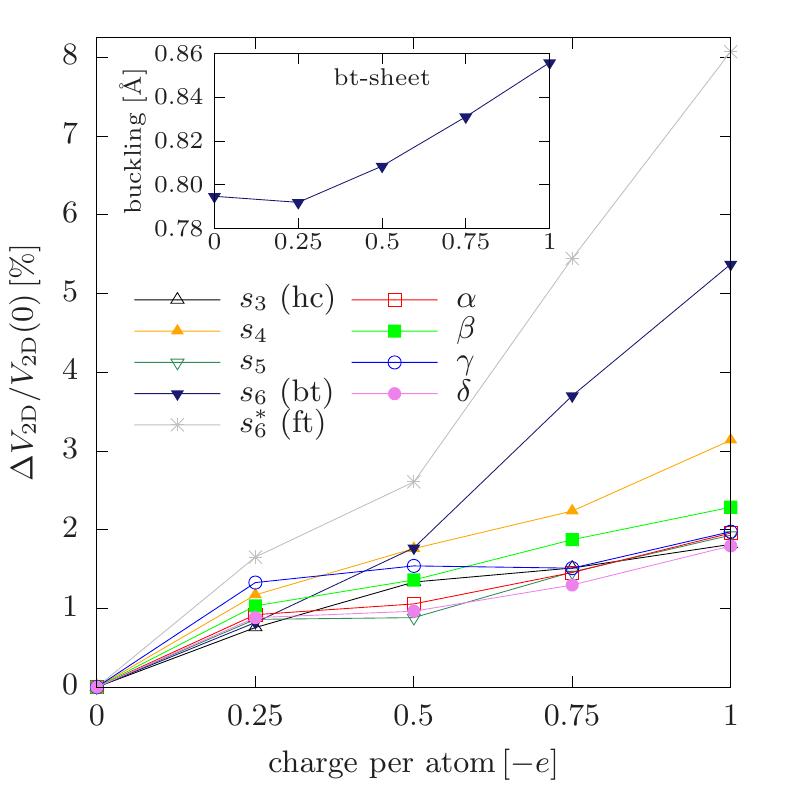}
  \caption{(Color online). Expansion of the unit cell 2D volume versus excess charge per atom. For each considered structure, the values are given relative to the 2D volume of the corresponding neutral structure. Inset: Buckling of the bt-sheet versus excess of negative charge per atom.}
\label{fig4}
\end{figure}

\begin{table*}
  \caption{Structural properties of the neutral structures shown in Fig.~\ref{fig1}. In the table, $V_{2\textnormal{D}}$ represents the 2D volume per atom and $N$ the number of atoms in the conventional unit cell of dimensions $a$, $b$. For each structure, its binding energy, $E_{\textnormal{b}}$, is also provided.} 
  \centering
  \begin{tabular}{|lllllllllll|}
    \hline
    & $\alpha$& $\beta$& $\gamma$& $\delta$& $\varepsilon$& $s_3$ (hc)& $s_4$& $s_5$& $s_6$ (bt)& $s_6^*$ (ft)\\
    \hline\hline
    plane group & \textit{p}6\textit{mm} & \textit{cmm} & \textit{pmm} & \textit{cmm} & \textit{cmm} & \textit{p}6\textit{mm} & \textit{pmm} & \textit{p}6 & \textit{pmm} & \textit{p}6\textit{mm} \\
    $a\, [\textnormal{\AA}]$& $5.0494$& $11.7674$& $5.0628$& $8.4017$& $10.1386$& $2.9098$& $3.3528$& $4.4576$& $2.8636$& $1.6985$\\
    $b\, [\textnormal{\AA}]$& $5.0494$& $2.9291$& $2.9120$& $2.9012$& $5.8562$& $2.9098$& $2.8729$& $4.4576$& $1.6247$& $1.6985$\\
    buckling& $0$& $0$& $0$& $0$& $0.1861$& $0$& $0$& $0$& $0.7948$& $0$\\
    $V_{2\textnormal{D}}\,[\textnormal{\AA}^2/\textnormal{atom}]$ & $2.76$ & $2.87$ & $2.95$ & $3.05$ & $2.70$ & $3.67$ & $3.21$ & $2.87$ & $2.33$ & $2.50$\\
    $N$& 8& 12& 5& 8& 22& 2& 3& 6& 2& 1\\
    hole density& $1/9$& $1/7$& $1/6$& $1/5$& $1/12$& $1/3$& $1/4$& $1/7$& $0$& $0$\\
    $E_{\textnormal{b}}\, [\textnormal{eV}/\textnormal{atom}]$& $6.6050$& $6.5791$& $6.5333$& $6.5413$& $6.5498$& $5.6419$& $6.3749$& $6.5723$& $6.5193$& $6.2976$\\
    \hline
  \end{tabular}
  \label{tab1}
\end{table*}

To understand the results presented in Fig.~\ref{fig2}, we propose a simple model in which we decompose the binding energy of 2D boron crystals into contributions of energies of the constituent atoms that have different coordination numbers. For that purpose, we express the binding energy of each structure as:
\[
E_{b}\left(n_3,n_4,n_5,n_6,q\right)=\frac{1}{N}\sum_{i=3}^6n_ie_i(q),
\]
where $n_i$ and $e_i$ are the number of boron atoms in the unit cell with $i$ nearest neighbors and their energy, respectively, and $N$ is the total number of atoms per unit cell. The individual energy contributions, $e_i$, are found from separate computations for structures with only one type of coordination number $i$. The idea is to have $E_\textnormal{b}\left(n_i,q\right)$ for several structures and compare the results with first principles calculations. The dependence of the $e_i$ energies on charge $q$ is shown in Fig.~\ref{fig3}a, whereas the comparison between the predicted and calculated binding energies for each considered charge is shown in Fig.~\ref{fig3}b. As can be seen from Fig.~\ref{fig3}b this model is too simple to predict the binding energies of 2D boron structures, i.e., the difference between the estimated and computed binding energies is for each neutral structure quite large and changes with the excess of negative charge. This might imply that the interactions between an atom and its second and more distant nearest neighbors should be also included in order to describe correctly the binding energy of neutral and charged structures. From Fig.~\ref{fig3}a, we can learn, however, that the $s_4$, $s_5$, and $s_6$ (bt-sheet) are the most energetically favorable structures (for neutral and negative charges) among those considered to build the model. Therefore, despite the limitations, based on our model we may conclude that the most stable neutral and negatively charged flat boron sheets should be those that have the largest number of four- and five-coordinated B atoms, and this is actually the case of the experimentally obtained $\delta$-sheet. On the other hand, the $s_3$ structure (hc-sheet) is the least favorable -- it is even worse than the ft-sheet for all the considered cases (with the exception of $q=-1.00\ e/\textnormal{atom}$). This means that boron layers that have tree-coordinated atoms should be the least stable.

\subsection{Structural properties}

In Tab.~\ref{tab1}, we summarize structural properties of the investigated structures. All of the structures described there, with the exception of $s_6$ (bt-sheet) and $\varepsilon$-sheet, are fully planar, i.e.~only $s_6$ and $\varepsilon$ exhibit periodic vertical displacements (buckling) of the atoms. Moreover, the $s_6$-sheet does not have 6-fold rotation point -- this symmetry is broken due to crystal deformation of the flat triangular lattice.

It is interesting to mention that boron will not adopt the honeycomb structure even if charged with negative charges as high as $-1.00\, e/\textnormal{atom}$. This is clear not only from the perspective of the binding energy for the hc-sheet that is always smaller than that of the, e.g., bt-sheet (see Fig.~\ref{fig2}), but also the hc-sheet will undergo distortions if the symmetry constrain is removed. Moreover, the 2D bulk modulus of the highly charged hc-sheet is $81\, \textnormal{N}/\textnormal{m}$ ($74.1\, \textnormal{N}/\textnormal{m}$, for the neutral structure), what is closer in value to that of neutral silicene ($72.00\, \textnormal{N}/\textnormal{m}$, taken from Ref.~\cite{dzade2010}) than that of neutral graphene ($211.8\, \textnormal{N}/\textnormal{m}$). 

Finally, in Fig.~\ref{fig4}, we show the expansion of the unit cell 2D volume versus excess of negative charge. From that figure we see that the largest expansion with increasing negative charge occurs for the flat-triangular sheet (followed by the bt-sheet), whereas the variation in 2D volume is much smaller for structures with hexagonal holes. Similarly, in the case of graphene, negative charging has little effect on its lattice constant, since the excess of electrons mostly go away from the structure \cite{topsakal2011}.

\section{Conclusions}
The performed studies reveal that there is a clear preference for 2D boron structures with very small or very high negative charge per atom. Structures with intermediate charges are energetically not favorable. Our analysis also suggests that under electron rich conditions there should be a clear preference for the formation of the $\delta$-sheet, since this structure exhibits the highest binding energy and is composed purely of four- and five-coordinated B atoms. Any static charging seems to be detrimental in terms of biding energy for the bt-sheet, however, this structure undergoes the second highest (after the ft-sheet) 2D volume expansion from all the considered boron sheets, what may be helpful in matching to the size of the unit cell of the metallic substrate.

\section*{Acknowledgements}
The authors gratefully acknowledge the support of the National Research Council (NCN) through the grant UMO-2013/11/B/ST3/04273. Numerical calculations were performed at ICM at the University of Warsaw under grant No.~G62-8.

\bibliographystyle{elsarticle-num}
\bibliography{v5.bib}

\begin{thebibliography}{10}
\expandafter\ifx\csname url\endcsname\relax
  \def\url#1{\texttt{#1}}\fi
\expandafter\ifx\csname urlprefix\endcsname\relax\def\urlprefix{URL }\fi
\expandafter\ifx\csname href\endcsname\relax
  \def\href#1#2{#2} \def\path#1{#1}\fi

\bibitem{li2015}
X.-B. Li, S.-Y. Xie, H.~Zheng, W.~Q. Tian, H.-B. Sun, Boron based
  two-dimensional crystals: theoretical design{,} realization proposal and
  applications, Nanoscale 7 (2015) 18863.
\newblock \href {http://dx.doi.org/10.1039/C5NR04359J}
  {\path{doi:10.1039/C5NR04359J}}.

\bibitem{evans2005}
M.~H. Evans, J.~D. Joannopoulos, S.~T. Pantelides, Electronic and mechanical
  properties of planar and tubular boron structures, Physical Review B 72
  (2005) 045434.
\newblock \href {http://dx.doi.org/10.1103/PhysRevB.72.045434}
  {\path{doi:10.1103/PhysRevB.72.045434}}.

\bibitem{kunstmann2006}
J.~Kunstmann, A.~Quandt, Broad boron sheets and boron nanotubes: An \textit{ab
  initio} study of structural, electronic, and mechanical properties, Physical
  Review B 74 (2006) 035413.
\newblock \href {http://dx.doi.org/10.1103/PhysRevB.74.035413}
  {\path{doi:10.1103/PhysRevB.74.035413}}.

\bibitem{cabria2006}
I.~Cabria, M.~J. L{\'o}pez, J.~A. Alonso, Density functional calculations of
  hydrogen adsorption on boron nanotubes and boron sheets, Nanotechnology 17
  (2006) 778.
\newblock \href {http://dx.doi.org/10.1088/0957-4484/17/3/027}
  {\path{doi:10.1088/0957-4484/17/3/027}}.

\bibitem{tang2007}
H.~Tang, S.~Ismail-Beigi, Novel precursors for boron nanotubes: The competition
  of two-center and three-center bonding in boron sheets, Physical Review
  Letters 99 (2007) 115501.
\newblock \href {http://dx.doi.org/10.1103/PhysRevLett.99.115501}
  {\path{doi:10.1103/PhysRevLett.99.115501}}.

\bibitem{penev2012}
E.~S. Penev, S.~Bhowmick, A.~Sadrzadeh, B.~I. Yakobson, Polymorphism of
  two-dimensional boron, Nano Letters 12~(5) (2012) 2441--2445.
\newblock \href {http://dx.doi.org/10.1021/nl3004754}
  {\path{doi:10.1021/nl3004754}}.

\bibitem{yu2012}
X.~Yu, L.~Li, X.-W. Xu, C.-C. Tang, Prediction of two-dimensional boron sheets
  by particle swarm optimization algorithm, The Journal of Physical Chemistry C
  116~(37) (2012) 20075--20079.
\newblock \href {http://dx.doi.org/10.1021/jp305545z}
  {\path{doi:10.1021/jp305545z}}.

\bibitem{wu2012}
X.~Wu, J.~Dai, Y.~Zhao, Z.~Zhuo, J.~Yang, X.~C. Zeng, Two-dimensional boron
  monolayer sheets, ACS Nano 6~(8) (2012) 7443--7453.
\newblock \href {http://dx.doi.org/10.1021/nn302696v}
  {\path{doi:10.1021/nn302696v}}.

\bibitem{lau2007}
K.~C. Lau, R.~Pandey, Stability and electronic properties of
  atomistically-engineered 2d boron sheets, The Journal of Physical Chemistry C
  111~(7) (2007) 2906--2912.
\newblock \href {http://dx.doi.org/10.1021/jp066719w}
  {\path{doi:10.1021/jp066719w}}.

\bibitem{zhou2014}
X.-F. Zhou, X.~Dong, A.~R. Oganov, Q.~Zhu, Y.~Tian, H.-T. Wang, Semimetallic
  two-dimensional boron allotrope with massless dirac fermions, Physical Review
  Letters 112 (2014) 085502.
\newblock \href {http://dx.doi.org/10.1103/PhysRevLett.112.085502}
  {\path{doi:10.1103/PhysRevLett.112.085502}}.

\bibitem{ma2016}
F.~Ma, Y.~Jiao, G.~Gao, Y.~Gu, A.~Bilic, Z.~Chen, A.~Du, Graphene-like
  two-dimensional ionic boron with double dirac cones at ambient condition,
  Nano Letters 16~(5) (2016) 3022--3028.
\newblock \href {http://dx.doi.org/10.1021/acs.nanolett.5b05292}
  {\path{doi:10.1021/acs.nanolett.5b05292}}.

\bibitem{ngsz2007}
N.~Gonzalez~Szwacki, Boron fullerenes: A first-principles study, Nanoscale
  Research Letters 3~(2) (2007) 49.
\newblock \href {http://dx.doi.org/10.1007/s11671-007-9113-1}
  {\path{doi:10.1007/s11671-007-9113-1}}.

\bibitem{tai2015}
G.~Tai, T.~Hu, Y.~Zhou, X.~Wang, J.~Kong, T.~Zeng, Y.~You, Q.~Wang, Synthesis
  of atomically thin boron films on copper foils, Angewandte Chemie
  International Edition 54 (2015) 15473.
\newblock \href {http://dx.doi.org/10.1002/anie.201509285}
  {\path{doi:10.1002/anie.201509285}}.

\bibitem{mannix2015}
A.~J. Mannix, X.-F. Zhou, B.~Kiraly, J.~D. Wood, D.~Alducin, B.~D. Myers,
  X.~Liu, B.~L. Fisher, U.~Santiago, J.~R. Guest, M.~J. Yacaman, A.~Ponce,
  A.~R. Oganov, M.~C. Hersam, N.~P. Guisinger, Synthesis of borophenes:
  Anisotropic, two-dimensional boron polymorphs, Science 350 (2015) 1513.
\newblock \href {http://dx.doi.org/10.1126/science.aad1080}
  {\path{doi:10.1126/science.aad1080}}.

\bibitem{zhang2016_2}
Z.~Zhang, A.~J. Mannix, Z.~Hu, B.~Kiraly, N.~P. Guisinger, M.~C. Hersam, B.~I.
  Yakobson, Substrate-induced nanoscale undulations of borophene on silver,
  Nano Letters 16~(10) (2016) 6622--6627.
\newblock \href {http://dx.doi.org/10.1021/acs.nanolett.6b03349}
  {\path{doi:10.1021/acs.nanolett.6b03349}}.

\bibitem{feng2016}
B.~Feng, J.~Zhang, Q.~Zhong, W.~Li, S.~Li, H.~Li, P.~Cheng, S.~Meng, L.~Chen,
  K.~Wu, Experimental realization of two-dimensional boron sheets, Nature
  Chemistry 8 (2016) 563.
\newblock \href {http://dx.doi.org/10.1038/nchem.2491}
  {\path{doi:10.1038/nchem.2491}}.

\bibitem{zhong2017}
Q.~Zhong, J.~Zhang, P.~Cheng, B.~Feng, W.~Li, S.~Sheng, H.~Li, S.~Meng,
  L.~Chen, K.~Wu, Metastable phases of {2D} boron sheets on {Ag(111)}, Journal
  of Physics: Condensed Matter 29~(9) (2017) 095002.
\newblock \href {http://dx.doi.org/10.1088/1361-648X/aa5165}
  {\path{doi:10.1088/1361-648X/aa5165}}.

\bibitem{zhang2016}
Z.~Zhang, E.~S. Penev, B.~I. Yakobson, Two-dimensional materials: Polyphony in
  b flat, Nature Chemistry 8 (2016) 525.
\newblock \href {http://dx.doi.org/10.1038/nchem.2521}
  {\path{doi:10.1038/nchem.2521}}.

\bibitem{qe2009}
P.~Giannozzi, S.~Baroni, N.~Bonini, M.~Calandra, R.~Car, C.~Cavazzoni,
  D.~Ceresoli, G.~L. Chiarotti, M.~Cococcioni, I.~Dabo, A.~{Dal Corso},
  S.~de~Gironcoli, S.~Fabris, G.~Fratesi, R.~Gebauer, U.~Gerstmann,
  C.~Gougoussis, A.~Kokalj, M.~Lazzeri, L.~Martin-Samos, N.~Marzari, F.~Mauri,
  R.~Mazzarello, S.~Paolini, A.~Pasquarello, L.~Paulatto, C.~Sbraccia,
  S.~Scandolo, G.~Sclauzero, A.~P. Seitsonen, A.~Smogunov, P.~Umari, R.~M.
  Wentzcovitch, \textsc{Quantum ESPRESSO}: a modular and open-source software
  project for quantum simulations of materials, Journal of Physics: Condensed
  Matter 21 (2009) 395502.
\newblock \href {http://dx.doi.org/10.1088/0953-8984/21/39/395502}
  {\path{doi:10.1088/0953-8984/21/39/395502}}.

\bibitem{VESTA}
K.~Momma, F.~Izumi, {\it VESTA3} for three-dimensional visualization of
  crystal, volumetric and morphology data, Journal of Applied Crystallography
  44 (2011) 1272.
\newblock \href {http://dx.doi.org/10.1107/S0021889811038970}
  {\path{doi:10.1107/S0021889811038970}}.

\bibitem{findsym}
H.~T. Stokes, D.~M. Hatch, {FINDSYM}: program for identifying the space-group
  symmetry of a crystal, Journal of Applied Crystallography 38 (2005) 237.
\newblock \href {http://dx.doi.org/10.1107/S0021889804031528}
  {\path{doi:10.1107/S0021889804031528}}.

\bibitem{ozdogan2010}
C.~\"{O}zdo\u{g}an, S.~Mukhopadhyay, W.~Hayami, Z.~B. G\"{u}ven\c{c},
  R.~Pandey, I.~Boustani, The unusually stable {B$_{100}$} fullerene,
  structural transitions in boron nanostructures, and a comparative study of
  $\alpha$- and $\gamma$-boron and sheets, The Journal of Physical Chemistry C
  114~(10) (2010) 4362--4375.
\newblock \href {http://dx.doi.org/10.1021/jp911641u}
  {\path{doi:10.1021/jp911641u}}.

\bibitem{amsler2013}
M.~Amsler, S.~Botti, M.~A.~L. Marques, S.~Goedecker, Conducting boron sheets
  formed by the reconstruction of the $\ensuremath{\alpha}$-boron (111)
  surface, Physical Review Letters 111 (2013) 136101.
\newblock \href {http://dx.doi.org/10.1103/PhysRevLett.111.136101}
  {\path{doi:10.1103/PhysRevLett.111.136101}}.

\bibitem{liu2013}
H.~Liu, J.~Gao, J.~Zhao, From boron cluster to two-dimensional boron sheet on
  cu(111) surface: Growth mechanism and hole formation 3 (2013) 3238, article.
\newblock \href {http://dx.doi.org/10.1038/srep03238}
  {\path{doi:10.1038/srep03238}}.

\bibitem{liu2013_2}
Y.~Liu, E.~S. Penev, B.~I. Yakobson, Probing the synthesis of two-dimensional
  boron by first-principles computations, Angewandte Chemie International
  Edition 125~(11) (2013) 3238--3241.
\newblock \href {http://dx.doi.org/10.1002/ange.201207972}
  {\path{doi:10.1002/ange.201207972}}.

\bibitem{zhang2015}
Z.~Zhang, Y.~Yang, G.~Gao, B.~I. Yakobson, Two-dimensional boron monolayers
  mediated by metal substrates, Angewandte Chemie International Edition 54~(44)
  (2015) 13022--13026.
\newblock \href {http://dx.doi.org/10.1002/anie.201505425}
  {\path{doi:10.1002/anie.201505425}}.

\bibitem{topsakal2011}
M.~Topsakal, S.~Ciraci, Static charging of graphene and graphite slabs, Applied
  Physics Letters 98~(13) (2011) 131908.
\newblock \href {http://dx.doi.org/10.1063/1.3573806}
  {\path{doi:10.1063/1.3573806}}.

\bibitem{dzade2010}
N.~Y. Dzade, K.~O. Obodo, S.~K. Adjokatse, A.~C. Ashu, E.~Amankwah, C.~D.
  Atiso, A.~A. Bello, E.~Igumbor, S.~B. Nzabarinda, J.~T. Obodo, A.~O. Ogbuu,
  O.~E. Femi, J.~O. Udeigwe, U.~V. Waghmare, Silicene and transition metal
  based materials: prediction of a two-dimensional piezomagnet, Journal of
  Physics: Condensed Matter 22 (2010) 375502.
\newblock \href {http://dx.doi.org/10.1088/0953-8984/22/37/375502}
  {\path{doi:10.1088/0953-8984/22/37/375502}}.

\end{thebibliography}

\end{document}